\documentclass{aa}
\usepackage{graphicx}
\usepackage{txfonts}
\usepackage{epstopdf}
\usepackage{natbib}
\usepackage{textcomp}			
\usepackage[colorlinks=true,citecolor=blue]{hyperref}
\bibpunct{(}{)}{,}{a}{}{;}

\begin{document}
\title{STAR CLUSTERS IN THE MAGELLANIC CLOUDS - II. AGE-DATING, CLASSIFICATION AND SPATIO-TEMPORAL DISTRIBUTION OF THE SMC CLUSTERS }
\author{P. K. Nayak   \inst{1}   \and
        A. Subramaniam   \inst{1} \and
        S. Choudhury \inst{2} \and
        Ram Sagar    \inst{1}   
        }
\institute{Indian Institute of Astrophysics, Koramangala II Block, Bangalore-560034, India. \\
	              \email{prasanta@iiap.res.in, purni@iiap.res.in}
	              \and
	              Yonsei University Observatory, 120-749, Seoul, Republic of Korea. \\ }
\abstract  
{}
{ We aim to estimate the age and reddening parameters of already identified star clusters within the Small Magellanic Cloud (SMC) in a consistent way using available photometric data, classify them based on their mass and strength, and study their spatio-temporal distribution.
 }
{We have used a semi-automated quantitative method, developed in the first paper of this series (Paper I), to estimate the cluster parameters using the {{\it V}} and {{\it I}} band photometric data from the Optical Gravitational Lensing Experiment (OGLE) III survey.  
}
{We estimated  parameters of 179 star clusters (17 are newly parameterised) and classified them into 4 groups. 
We present an on-line catalog of parameters as well as cleaned and isochrone-fitted Color Magnitude Diagrams (CMDs) of 179 clusters.
  
We compiled age information of 468 clusters by combining previous studies with our catalog, to study their spatio-temporal distribution. Most of the clusters located in the southern part of the SMC are in the age range 600 Myr - 1.25 Gyr, whereas, the clusters younger than 100 Myr are mostly found in the northern SMC, with the central SMC showing continuous cluster formation. The peak of the cluster age distribution is identified at 130 $\pm$ 35 Myr, very similar to the Large Magellanic Cloud (LMC) in Paper I.
}
{We suggest that the burst of cluster formation at 130 Myr is due to the most recent LMC-SMC interaction.  
 90 $\%$ of the studied sample is found to have mass $<$ 1700 $M_{\odot}$, suggesting that the SMC is dominated by low mass clusters.  There is a tentative evidence for compact clusters in the LMC when compared to those in the Galaxy and the SMC.
 A progressive shifting of cluster location from the South to North of the SMC is identified in last $\sim$ 600 Myr. 
The details of spatio-temporal distribution of clusters presented in two videos in this study can be used as a tool to constrain details of the recent LMC-SMC interactions. 
}

 
\authorrunning{P. K. Nayak et al.}
\titlerunning{Age-Dating, Classsification and Spatio-Temporal distribution of the SMC clusters}
\maketitle

\section{Introduction}

The Small Magellanic Cloud (SMC), located at a distance (D) of $\sim$ 60 kpc is a nearby dwarf galaxy to the Milky Way (MW) beyond the Large Magellanic Cloud (LMC, D $\sim$ 50 kpc). The SMC is classified as an irregular galaxy with a less pronounced bar and is known to be tidally disturbed because of its ongoing interaction with the LMC and the MW \citep{besla2010,besla2012}. The LMC and the SMC are enclosed within an extended body of diffuse HI gas that extends out to many tens of degrees across the sky, forming the Magellanic Stream and the Leading Arm \citep{wann1972,math1974,putman2003,nid2010}. 
These features provide ample evidence of the MW-LMC-SMC interactions \citep{besla2010,besla2012,diaz2011,diaz2012}. 

There have been several advances in understanding the interaction between these three galaxies over the last decade. \cite{diaz2012} suggested that Magellanic Clouds (MCs) have undergone at least two pericentric passages about the MW during a $\sim$ 2 Gyr bound association. On the other hand, the recent high precision measurement of proper motion of the MCs using the Hubble Space Telescope (HST) data suggests that either the MCs are undergoing their first passage near to the MW \citep{kalli2013} or they are orbiting with a long period ($>$ 6 Gyr) around the Galaxy \citep{besla2010}. The proper motion study of the LMC using HST data suggested that the MCs have just passed their pericentre (45 kpc from the Galactic centre) with apocentre to pericentre ratio of 2.5:1 with an orbital period of 1.5 Gyr \citep{kalli2006a}. \citet{diaz2011,diaz2012} suggested that the SMC became a strongly interacting binary pair with the LMC only  recently, suffering two strong tidal interactions $\sim$ 2 Gyr ago and $\sim$ 250 Myr ago.  According to \citet{besla2012}, the SMC made close passages around the LMC at around 900 Myr and 100 Myr ago. These strong interactions between the MCs not only pulled out gas from the disc of the SMC, but also stars. The Gaia DR1 data revealed stellar tidal tails around both the Clouds and an almost continuous stellar bridge \citep{belo2017} connecting the two clouds. A significant number of Miras were found in the East of the LMC by \cite{deason2017} using the first data released from Gaia mission and they inferred that these are likely to be stripped away from the SMC due to interaction with the LMC. 

The close encounter between the MCs can also trigger star formation in both the clouds. Using Magellanic Clouds Photometric Survey (MCPS, \cite{zaritsky2002,zaritsky2004}) data, \cite{harris2004} showed that the burst of star formation happened at ages of 2.5, 0.4 and 0.06 Gyr in the SMC.  
The burst timescales more or less coincide with the two past encounters between the MCs. The peaks at 2.5 and 0.4 Gyr also coincide with the star formation peaks in the LMC \citep{harris2009}. Any such triggered star formation can also lead to the formation of star clusters within a galaxy. Also, any propagation of cluster formation within a galaxy can indicate interaction details. 
Hence, it will be useful to investigate the cluster formation history in the SMC along with their spatio-temporal distribution and correlate the peaks of cluster formation with the epoch of interaction.

A number of previous studies have been carried out to identify star clusters within the SMC. The recent and the most extensive study is presented by \cite{bica2008} (hereafter B08), where the authors listed $\sim$ 600 star clusters in the SMC with their central co-ordinate, values of major and minor diameters, and position angle. However, the ages, reddening and mass of the clusters are not available in their catalog. Using the MCPS data, \cite{glatt2010} (hereafter G10) estimated ages and reddening of 324 objects in the SMC, which include clusters and associations from the catalog of B08. G10 found the age distribution of the clusters to have peaks at 160 Myr and 630 Myr, and they suggested that the interaction between the MCs resulted in the formation of these peaks. \cite{PU99_smc} (hereafter PU99) provided  the age information of 93 well-populated SMC clusters using Optical Gravitational Lensing Experiment (OGLE) II survey data \citep{udalski1998}. \cite{chiosi2006} presented the ages of 311 clusters younger than 1 Gyr. The authors used two sets of data for their analysis: OGLE II for the SMC disk and the data obtained from the ESO 2.2 m telescope for the region around NGC 269, located in the South-East end of the disk. These authors found that the age distribution of clusters showed  an enhancement between 15 Myr to 90 Myr.

Studies by \cite{pia2005,pia2007b,pia2007a,pia2007c,pia2008,pia2011a,pia2011b,maia2012} made use of deep Washington photometric data concerned primarily intermediate and old clusters in the SMC. 
\cite{mighell1998,glatt2008,girardi2013} studied intermediate and old clusters using photometric data obtained from the HST.  Age, metal abundance and positional data of 12 star clusters were presented by \cite{crowl2001}. 
Ages of 15 intermediate to old star clusters were determined \cite{parisi2014}. 
\cite{dias2014} derived age, metallicity, reddening and distance for star clusters in the SMC west halo.
Recently \cite{pia2015} studied 51 star clusters in the eastern outskirt of the SMC and in the bridge region using the VMC survey. Based on the CMDs they defined 15 cataloged clusters to be possible non-genuine aggregates.

Despite several past studies, the number of SMC clusters with age information is only about 50 $\%$ of that listed by B08. Also, as mentioned by the authors, the list of detected clusters in the SMC is still incomplete (mostly restricted by poor detection limit). Therefore, to understand cluster formation history in detail and the effect of LMC-SMC-MW interactions on cluster formation, one has to determine the ages of already cataloged clusters, at the same time new clusters have to be identified using relatively deeper and large scale photometric data.

In this study, we have tried to estimate the age and reddening of the already identified star clusters by B08. The usual method of estimating the age of a cluster is by visual fitting of isochrones over the cluster CMD, which can produce a systematic error in the age estimation from cluster to cluster. At the same time, when cluster sample is large (more than hundred), it is a tedious job to fit isochrones to each cluster. Therefore, we have used the semi-automated quantitative method developed by \cite{nayak2016} (hereafter Paper I) to estimate the age and reddening. Using this method we also quantify the error in age.

The SMC was known to host primarily rich clusters. Nevertheless, recent studies suggested that it also contains poor clusters \citep{piatti2012} and hosts clusters with a wide range of masses \citep{hunter2003}. Using integrated colour in UBVR passbands and evolutionary models, \citet{hunter2003} estimated masses of 191 clusters in the SMC having a range between 10$^2$ to 10$^6$ M$_\odot$. \cite{konti1982} calculated masses of 20 clusters in the SMC using \citet{king1962} model and found that derived masses are about 10 times smaller than those in our Galaxy. Using archival HST snapshot data, \citet{mack2003} determined masses of 10 rich clusters from their surface brightness profiles. The estimated mass of those clusters range from 10$^{3.6}$ to 10$^{5.5}$ M$_\odot$. Recent study by \citet{maia2014} has provided masses of 29 young and intermediate clusters within a range of 300 to 3000 M$_\odot$ .

So far, there has not been any study to systematically classify the clusters based on their mass. \citet{searle} classified 61 star clusters in the MCs as type I to VII based on four colour photometry of integrated light. On the other hand, there are well known classification schemes for Galactic Open Clusters \citep{trump30} based on the degree of central concentration of stars, the range in luminosity of the members, the number of stars contained in the cluster and necessary conspicuous properties \citep{ruprecht66}. In this paper, we have tried to estimate the mass range of the SMC clusters and classify them based on their mass/richness. Estimation of the mass and age of cluster sample will help in understanding cluster formation, evaporation, and dynamical evolution of the cluster system.

Thus, the aims of this study are $\colon$ (1) to estimate age and reddening of already identified star clusters of the SMC in a consistent way using available photometric data and increase the sample of well-studied clusters (2) to classify the SMC clusters based on their mass/richness (3) to study the spatio-temporal distribution of the SMC clusters.

The rest of the paper is arranged as follows: In section 2 we have mentioned about the data used for this study, followed by analysis in section 3. Section 4 presents error estimation. Estimation of cluster mass and the classification scheme are described in section 5. Results of this study are presented in section 6, followed by the summary in section 7.

\section{Data}

We have used {\it V} and {\it I} photometric data from the OGLE III \citep{udalski2008_smc} survey to identify star clusters, which are listed in B08. The OGLE III observations were carried out at the Las Campanas Observatory,  Chile, between June 2001 and January 2008, with the 1.3-m Warsaw telescope equipped with second generation mosaic camera \citep{udalski2003} consisting of eight SITe 2048 $\times$ 4096 CCD detectors with pixel size of 15 $\mu$m, which corresponds to 0.$\arcsec$26. The field of view of the telescope is approximately 35 $\times$ 35 arcmin$^2$ on the sky. The survey covered a total area of about 14 square degrees in the sky around the SMC centre and produced a catalog of V and I magnitudes of about 6.2 million stars \citep{udalski2008_smc}. The completeness of the photometry is better than 75$\%$  in I band and 85$\%$ in V band for 20 mag for crowded regions.

In this study, we estimate the parameters (age, reddening, mass, richness) of star clusters, which can be affected by photometric incompleteness of the data. Photometric incompleteness will lead us to a wrong determination of cluster richness. A rich cluster located in the crowded field can appear as poor cluster due to photometric incompleteness. Luminosity function (LF) of a cluster will also be biased due photometric incompleteness and it will guide to a wrong estimation of mass function as well as mass of the clusters. The method used here to estimate age and reddening of clusters will also be affected by photometric incompleteness as well as photometric error. Therefore, we have considered stars having photometric errors $\le$ 0.15 mag in V and I bands with photometric completeness more than 90$\%$ even in the most crowded region to construct LF. We use the full data when it comes to plotting the colour magnitude diagrams (CMDs).

\section{Analysis}
\subsection{Cluster sample}

We adopted the most extensive SMC cluster catalog by B08 as a reference and identified 492 star clusters located well within the OGLE III observed region, mentioned in the last section. We define the radius of clusters to be \textonequarter(major + minor) diameter, the values of which are given in B08 catalog. The  estimated cluster radii are found to range from 0.$\arcmin$07 to 1.$\arcmin$70 on the sky, with physical sizes corresponding to a range of $\sim$ 1.22 to 29.6 pc. 
We extracted data of the cluster regions (stars within the cluster radius) from the OGLE III catalog along with a few arcmin field around them. 
These data are used for further analysis. Further details about the extraction are described in Paper I.

A star cluster is defined as a gravitationally bound system of a group of stars and can be observed as a density enhanced region with respect to its surrounding field region. The observed cluster region not only consists of cluster members but also has foreground and background field stars. The SMC hosts a good number of rich as well as poor clusters, located in a varied range of stellar density environments.  The fundamental features of a cluster which can be used to estimate the reddening, age and distance are the main-sequence (MS) and the location of MS turn-off in the CMD. So, field star removal is necessary to define the cluster sequence and for a better estimation of cluster parameters. 
We first identified those clusters, which are located in an environment with varying field star density. The estimated number of stars within the cluster radius is denoted as $n_c$. 
In order to estimate variation in the field star density, we chose four annular regions (each of equal area as the cluster region), of inner radii 0.$\arcmin$5, 1.$\arcmin$0, 1.$\arcmin$5 and 2.$\arcmin$0 larger than the cluster radius and counted the number of stars in each annular region. The number of field stars, which is an average of estimation from the four field regions ($n_f$) is considered to be contaminating the cluster region. Standard deviation ($\sigma_f$) about this average indicates the variation in the field star density.  The number of cluster members ($n_m$) or strength of the cluster is defined as $n_m = n_c - n_f$. Many clusters are found to have $\sigma_f$ $\sim$ $n_m$, suggesting that the cluster strength is similar to the fluctuation in the field star distribution. Then, we proceed to exclude clusters which have the following properties : \\
	(i) We estimated the fractional standard deviation as $\sigma_f/n_f$, to quantify the variation in field star counts. We excluded those clusters where variation in the field stars count is greater than or equal to 50$\%$ of the average count, i.e. $\sigma_f/n_f$ $\ge$ 0.5. With this criteria, we excluded 48 star clusters.\\
	(ii) The variation in the field star counts will propagate as an error when we estimate the strength of the cluster. The error associated with the estimation of $n_m$ can be defined as :\\
\begin{equation}
		e= |(n_c - (n_f - \sigma_f)) - (n_c - (n_f + \sigma_f))|,
\label{error_relation}
\end{equation}

which is basically the difference between the maximum and the minimum values of $n_m$, for a $\sigma_f$ deviation in the field star distribution. For crowded field regions, there is a possibility that $\sigma$$_f$ is high and so will be the value of $e$. In order
to remove clusters where the error itself is greater than the number of stars in the cluster, 
we calculated the fractional error as $e/n_m$ and excluded clusters with $e/n_m$ $\ge$ 1. Based on this criteria, we excluded 126 clusters from our sample.

The number of clusters remaining in the sample, after implementing the above two cut-off criteria is 337. Out of these clusters, 5 are relatively rich clusters ($n_m$ $>$ 400) and we have excluded them from our analysis as they are already well studied using better observational data. Thus, we proceeded with a sample of 332 clusters to decontaminate their cluster CMDs.
This exercise also points out that $\sim$ 30 $\%$ of the clusters are located in regions with significant variation in field star density. In Paper I, we found that about 20$\%$ of clusters are located in similar environments within the LMC.

 We constructed (V, V $-$ I) CMDs for cluster and field regions to compare and decontaminate the cluster region from field stars using a statistical process. For the details of the process, we direct the readers to section 3.2 of Paper I.

\subsection{Semi-automated quantitative method}

We have adopted the semi-automated quantitative method developed in Paper I, to estimate cluster parameters accurately and consistently. We also quantify error using this method. The method is applied to all the 332 clusters to estimate reddening and age.

The primary steps involved in the method are to :\\
(a) Identify the MS in the cleaned cluster CMD and construct the MS luminosity function (MSLF).\\
(b) Identify the MS Turn-off (MSTO) from the MSLF and estimate the corresponding apparent magnitude and colour.\\
(c) Estimate the reddening from the (V$-$I) colour of the MSTO.\\
(d) Estimate the absolute magnitude of the MSTO after correcting for reddening and distance.\\
(e) Estimate the age using age-magnitude relation derived using \citet{marigo2008} (hereafter M08) isochrones.\\

The above steps are described in detail below.

		(a) We consider stars brighter than 21 mag in V and bluer than 0.5 mag in (V $-$ I) colour as the MS stars. To construct the MSLF, the magnitude axis is binned with a bin size of 0.2 mag. The brightest bin with a minimum number of stars ($\eta$) is identified as the bin corresponding to the MSTO. The mean V magnitude of the brightest bin is considered as turn-off V (V$_{TO}$) magnitude. The MSTO bin (which is likely to be the brightest bin of the MSLF) needs to be identified from the MSLF using statistically significant value of $\eta$ so that it excludes blue super giants. The value of $\eta$ will depend mainly on richness as well as age of the cluster.
Two clusters which are similar in richness, but with different age will have different MSTO bin with different $\eta$. The MSTO bin will be less populated for a younger cluster than the older one with similar richness. Two clusters with same age but a different number of cluster stars will also have different values of $\eta$ for their MSTO bin. So, the identified bin and the number of stars in the bin are dependent on the richness/age of the cluster. 
Here, the only known parameter is the richness (total number of cluster stars) of the cluster, as we are yet to estimate their age. Therefore, we have grouped the clusters according to their strength, similar to Paper I, and considered similar $\eta$ value  corresponding to each group to identify the MSTO bin. In the next section, we have briefly described the calibration procedure of $\eta$. We refer to section 3.4 of Paper I for more details. 
After classifying the group number, strength of clusters and the $\eta$ value corresponding to each group are tabulated in Table \ref{number_criteria}.

		(b) Once we have identified the V$_{TO}$, the next task is to estimate the colour of the MSTO. The colour of the MSTO can be identified as the peak in colour distribution near the MSTO.
To estimate the peak colour of the MSTO,  a strip parallel to the colour axis with a width of 0.6 mag about V$_{TO}$ is considered (V$_T$$_O$ + 0.4 mag to V$_T$$_O$ - 0.2 mag). This is to ensure that we have a statistically significant number of stars near the MSTO. For the clusters with $n_m\le$ 100, a width of 0.8 mag is considered (given by V$_T$$_O$ + 0.6 mag to V$_T$$_O$ $-$ 0.2 mag). The choice for width of the strip does not affect the position of the peak colour, as the isochrones for younger ages are almost vertical to the colour axis near the MSTO. This strip is binned in colour with a bin size of 0.1 mag to estimate the distribution of stars along the colour axis. The distribution is found to have a unique peak (in most of the cases) with asymmetric wings. The mean colour of the bin corresponding to this unique peak is chosen as the apparent color, (V $-$ I)$_{app}$, of the MSTO.

		(c) The reddening of the cluster is defined as the difference between the apparent and absolute colour of the MSTO. 
To begin with, we have adopted A$_V$ = 0.46 mag \citep{zaritsky2002} 
and distance modulus of DM = 18.90 mag for the SMC \citep{storm2004}.
If M$_V$ is the absolute magnitude of the MSTO, then assuming a distance modulus and an average value of extinction (A$_V$) for the cluster, the apparent magnitude (V$_{TO}$) is related to M$_V$ as:
\begin{equation}
		M_V = V_{TO} - DM - A_V,
\label{mv_vto_relation}
\end{equation}

The estimated value of M$_V$ is cross-matched with the absolute V magnitude of MSTOs from the isochrones table of M08 for a metallicity of Z = 0.004 
for the SMC. The (V $-$ I) colour corresponding to the closest match of absolute V magnitude of MSTO gives the true colour for the MSTO, (V $-$ I)$_0$. The reddening (E(V $-$ I)) for the cluster is then given as:
\begin{equation}
		E(V - I) = (V - I)_{app} - (V - I)_0. 
\label{evi_relation}
\end{equation}

  (d) The extinction for the cluster region is estimated as, A$_V$ = 2.48$\times$E(V $-$ I) \citep{niko2004}.
 The extinction corrected value of M$_V$ of the MSTO is then calculated again by using this value of A$_V$ in
Equation \ref{mv_vto_relation}. The values are found to be invariant even after a couple of iterations.
The method used here is similar to that adopted by \citet{indu2011}, for estimating the reddening
of field regions.

\begin{figure}
\includegraphics[width=\columnwidth]{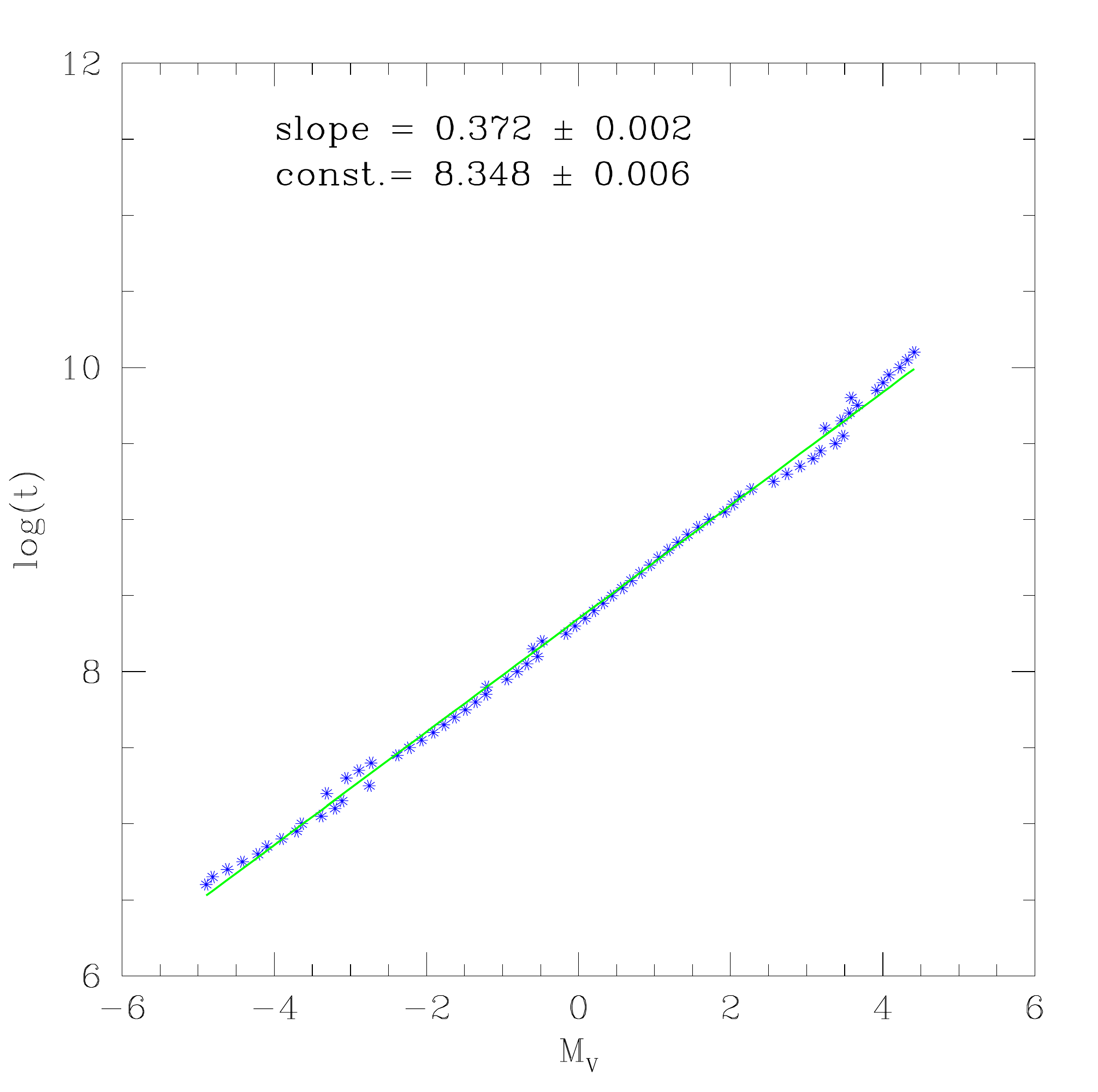}
\caption{The relation between the absolute turn-off V magnitude (M$_V$) and age, within the range log(t) = 6.2 to 10.2 for M08 isochrones. A straight line (green) fitted through the points is also shown.}
\label{log(age)-Mv}
\end{figure}

		(e) Figure \ref{log(age)-Mv} shows relation between the absolute magnitude M$_V$ of the MSTO and their corresponding ages (log(t)) for M08 isochrones. The relation is found to be linear and is given as:
\begin{equation}
		log(t) = 0.372(\pm0.002)M_V + 8.348(\pm0.006). 
\label{logt_mv_relation}
\end{equation}
The extinction corrected M$_V$ derived in step (d), is used in the above relation to estimate the ages of the clusters.

 Once we estimated the parameters, we over plotted isochrones on each cluster CMD of corresponding age, after correcting for estimated reddening and extinction. We visually checked all the CMDs for any improper estimation of parameters. If required, we adjusted the parameters to improve the isochrone fit. As we have finalised the parameters after visual inspection, the method is termed as semi-automated quantitative method.

 This method primarily depends on the unique identification of the MSTO. As mentioned earlier, it is a function of age and richness of the cluster. Due to the photometric limit of OGLE III data (21 magnitude in V band), we have restricted the method to clusters with MSTO magnitude ($V_{TO}$) brighter than 19 mag for reliable estimation of clusters parameters. There are 189 clusters with $V_{TO} \le$ 19 mag. We visually checked the cluster CMDs with over plotted isochrones.  
 We found that the isochrones fitted very well for 62 clusters (32.8$\%$) and a small correction in age and/or reddening were required for 75 clusters (39.7$\%$). The remaining 52 clusters (27.5$\%$) have ambiguous cluster sequence. To identify the cluster sequence we decontaminated the cluster region for these 52 clusters with two more field regions and  over-plotted all decontaminated CMDs. The sources which have not been removed for at least two different decontamination processes were considered as a cluster member. Out of 52 clusters, 12 were found to show prominent cluster features, and required minor modification in age or/and reddening, to the automated estimation. Whereas, 40 clusters were found to have no clear feature in the CMD prohibiting any reliable estimation of parameters. Among these 40 clusters, four are found to be in common with the cluster candidates identified by \citet{pia2015}.

For star clusters in the LMC (Paper I), the semi-automated quantitative method worked well for more than 80 $\%$ of the total sample. The reason for reduced success rate in the case of the SMC is probably due to less number of cluster members (n$_m \le$ 30), resulting in sparsely populated CMDs and MS.

To increase the number of parameterised clusters, we also inspected the 143 cases (out of 337) with $V_{TO} >$ 19 mag. After
the visual inspection of their CMDs, 30 clusters are found to have reliable estimation (including a minor correction in estimated age and/or reddening with respect to automated estimation for a few). Thus, we were able to estimate the parameters of a total 179 clusters within the SMC.

\subsection{Calibration procedure of $\eta$}

The value of $\eta$ depends on richness and age of the cluster. Limited by the photometric depth,  OGLE III data is ideal to estimate the ages of mainly younger clusters (upto few hundred Myr) and $\eta$ will not depend much on this small age spread, as we have seen in Paper I. Therefore, it is necessary to fix the value of $\eta$ for different groups of clusters to estimate the parameters. After we grouped the clusters based upon their strength, we looked for the clusters from each group whose parameters are already estimated by G10. Then we estimated age and reddening of those clusters from a particular group for a range of $\eta$ values. We compared our age and reddening estimations with that of G10 for all $\eta$ values. We chose the $\eta$ value for which we found that deviations in the estimated parameters (age and reddening) are least and there are no systematic deviations with respect to G10's estimations. 
This is the way we calibrated the $\eta$ value for each group of clusters.

We also calibrated the $\eta$ value by generating synthetic CMDs and comparing it with observed CMDs. In this study, we found that ages of the SMC clusters peak at $\sim$100 Myr and reddening peaks between 0.10-0.20. We took age to be 100 Myr and reddening value as 0.15 to produce synthetic CMDs. In this analysis, we used Padova isochrone model \citep{marigo2008} and Salpeter$'$s mass function \citep{salpeter1955}. We produced synthetic CMDs by populating stars in the main sequence (MS) for observed ranges of V and I mag. We have also taken care of photometric incompleteness while generating synthetic CMDs. We calculated the number of stars present in the turn-off bin, which is nothing but the value of $\eta$, for different groups (I$-$V) of clusters. We have run the simulation for multiple iteration with different initial normalising star-counts. We have also run the simulation for another two age values (200 and 300 Myr). We found that the $\eta$ values estimated from the above two methods match very closely and are tabulated in Table 1.

\begin{table}
\centering
\caption{ Grouping and classification of clusters based on their richness ($n_m$) and mass range (M$_c$):}
\label{number_criteria}
\resizebox{89mm}{!}{
\begin{tabular}{ccccccc}
\hline
Group No. & Range of $n_m$  & $\eta$ & $\eta_{simulated}$ & N$_{total}$  & Mass range (M$_\odot$) & Classification \\ \hline
 I        & 6$<n_m\le$30    & 2      &  1.23              & 94           & $<$ 800                & very poor  \\  \hline       
 II       & 30$<n_m\le$100  & 3      &  2.88              & 69           & 800 - 1700             & poor \\ \hline            
 III      & 100$<n_m\le$200 & 5      &  6.16              & 13           & 1700 - 3500            & moderate\\ \hline          
 IV       & 200$<n_m\le$300 & 10     & 10.27              & 2            & 3500 - 5000            & moderate\\ \hline
 V        & 300$<n_m\le$400 & 14     & 14.38              & 1            & $>$ 5000               & rich \\ \hline
\end{tabular}
}
\end{table}

\section{Error estimation}
		We have calculated the errors associated with the estimated age and reddening using the method of propagation of errors, adopted from Paper I. The error in estimating the reddening depends upon the photometric error and binning resolution along the colour axis, and this error will propagate to age estimation. The error associated with the age estimation depends on the errors in estimating extinction and absolute magnitude, and binning resolution along the magnitude axis. We have also considered the effect of distance spread in the SMC on age estimation. The SMC has a large range of line of sight depth of the SMC from 670 pc to 9.53 kpc (0.025 to 0.34 mag; \cite{smitha2009}). In our error analysis we considered the maximum depth of the SMC (9.53 kpc). As the reddening is estimated using the stars in the upper MS and the photometric error is very small ($\le$ 0.05) in the upper MS, the effect of photometric error in estimating reddening could be neglected. Thus, the error in the estimated reddening, E(V$-$I) is chosen to be same as the bin size, 0.1 magnitude. Errors in the estimation of extinction and age are given by the following relations $\colon$\\

$\sigma A_V$ = $ 2.48  \sqrt {\sigma{(V-I)}^2 + {(V-I)}_{bin}^2}$ \\

$\sigma$$M_V$ = $ \sqrt {\sigma V^2 + V_{bin}^2 + \sigma A_V^2 + \sigma (DM)^2}$ \\

$\sigma$(age) = constant x $\sigma$$M_V$ \\

where $V_{bin}$ \& ${(V-I)}_{bin}$ are half the bin sizes used for magnitude \& color binning,  $\sigma$$M_V$ is the error in absolute magnitude, $\sigma$$A_V$ is the error in the estimated extinction, $\sigma (DM)^2$ is the uncertainty in the distance modulus due to maximum line of sight depth in the SMC and $\sigma$(age) is the error in estimated age in (log(t)). The maximum error in estimated age is 0.25.

 The studies of intermediate and old SMC clusters by \citet{parisi2009} suggested that metallicity of the SMC clusters ranges from $-$0.60 to $-$1.30 dex with a mean of $-$0.96 dex. Therefore, we have also examined the effect of metallicity on estimated age. We derived M$_V$ vs log(age) relation for three different metallicities (Z = 0.001, 0.004, 0.008) and found that the variation in the slope and the y-intercept are in second and third decimal place respectively. The error in the age estimation varies from 0.24 to 0.26 in log scale due above mentioned metallicity range.
\\

\section{Estimation of mass range}
	We have divided the total cluster sample into five different groups (group I - group V) based on their strength ($n_m$) using the same criteria as in Paper I.
 In Table \ref{number_criteria}, the group numbers and corresponding range of cluster strength are listed in column 1 and column 2 respectively. Column 3 gives the list of $\eta$ values for different groups, which helps to identify turn-off magnitude in the cluster CMD. Cluster strength not only gives information about richness of the cluster but also indicates the mass of the cluster. 
We have tried to estimate the mass range for clusters corresponding to each group. Then, we classified the clusters based on their mass range using similar classification scheme as in Paper I.

	In our sample, we found that most of the clusters are younger than 300 Myr and the age distribution peaks at around 100 Myr. Hence, we assumed a typical age of 100 Myr and used the value of $\eta$ to estimate the mass (M$_c$) range of clusters. We constructed synthetic CMDs using M08 isochrones, for a mass range of 0.1 - 15.0 M$_\odot$. We assumed the mass function to be Salpeter$'$s mass function \citep{salpeter1955} and included observational errors. To reduce statistical fluctuation because of low $\eta$ value, we simulated the synthetic CMD for a large number of stars ($\sim$ 10$^6$) and scaled for the value of $\eta$ in the MSTO bin. The total mass of the cluster is estimated using the scaling factor. We found that the clusters in group I have M$_c$ $<$ 800 M$_\odot$ and we classify them as {\it very poor}. Clusters in group II have M$_c$ in the range $\sim$ 800 - 1700 M$_\odot$ and classified as {\it poor}. Group III and IV have been classified as {\it moderately rich} clusters with M$_c$ range  $\sim$ 1700 - 5000 M$_\odot$ and the clusters in the group V as {\it rich} clusters with M$_c$ $>$ 5000 M$_\odot$. In column 5 of table \ref{number_criteria}, mass range of clusters for different groups are listed and in column 6 the classification of clusters are noted. The total number of clusters (N$_{total}$) in each group are listed in column 4.

	The estimated mass ranges for different groups indicate that the SMC consists of clusters with a large mass range, similar to the LMC star clusters (Paper I). The classification of clusters based on mass range will help us to understand various properties, like formation and evolution of clusters, which depend on its mass. It will also help us in understanding the dissolution of star clusters in various groups in the SMC. 
	Our estimated mass ranges match well with that of \citet{maia2014} and \citet{hunter2003}. We have also compared the mass range of the SMC clusters with that of the open clusters in the Galaxy. We found that the clusters near solar neighbourhood have mass range \citep{lamers2005} similar to that in the SMC. \citet{pisku2008} studied 650 Galactic open clusters with mass range 50 M$_\odot$ to 10$^5$ M$_\odot$. Thus, SMC consists of clusters with a large mass range which is similar to Galactic open clusters. In our sample, about 50$\%$ of the clusters belong to the very poor group, suggesting the presence of a large fraction of very low mass clusters in the SMC. On the other hand, in Paper I we found that the LMC has $\sim$ 40$\%$ of very poor clusters. We also note that 90 $\%$ of the clusters are either in the poor or very poor class. All these clusters have mass $<$ 1700 M$_\odot$. 
	These suggest that the cluster population in the SMC is dominated by low mass clusters. This finding has implication to the cluster formation mechanism in the SMC.

\begin{figure}
\includegraphics[width=\columnwidth]{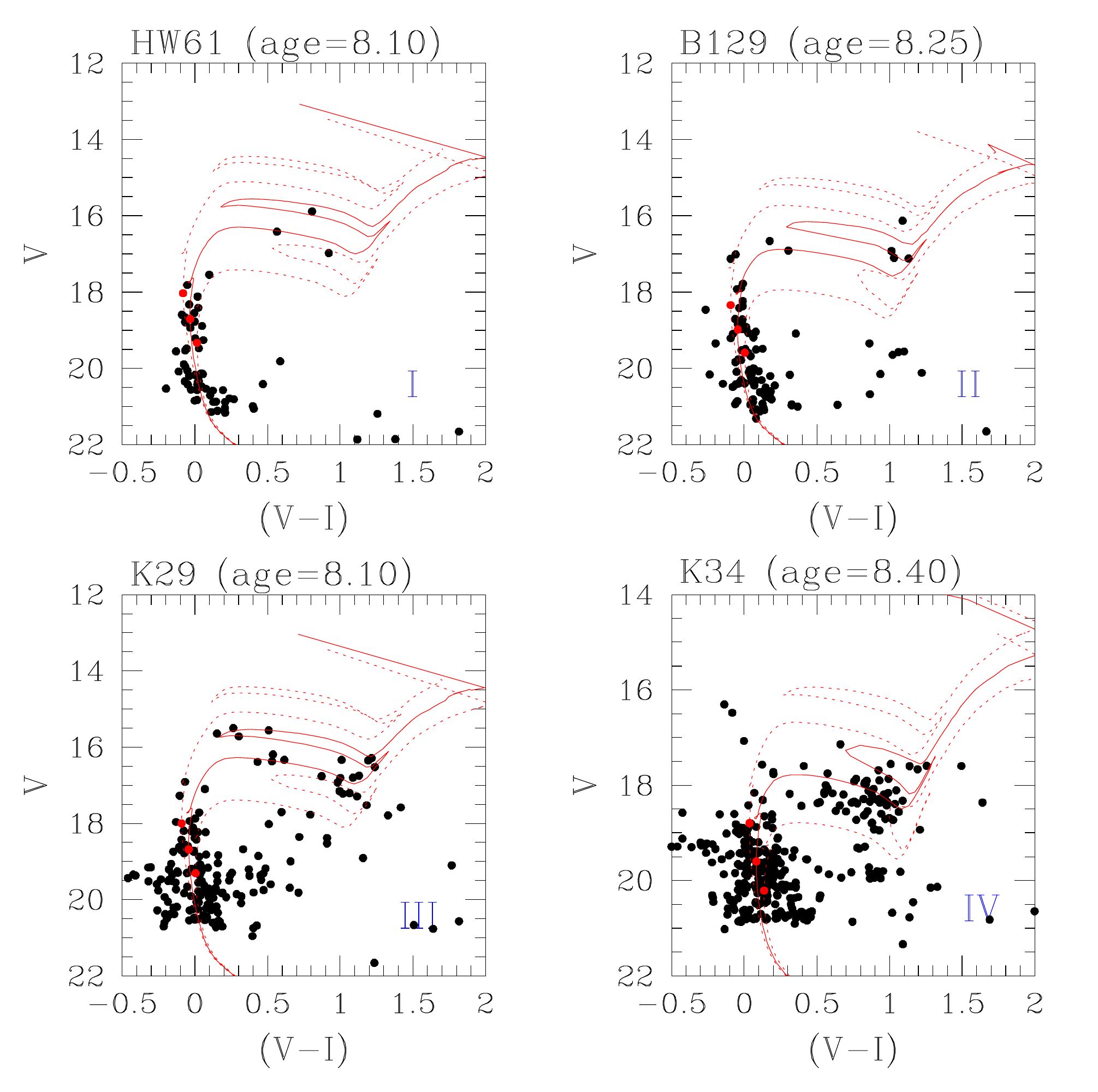} 
\caption{The plot shows CMDs of clusters from each group (I - IV). The cluster's name and age (log(t)) are also marked.}
\label{cmd}
\end{figure}

\begin{figure*}
\centering
\includegraphics[width=1.8\columnwidth]{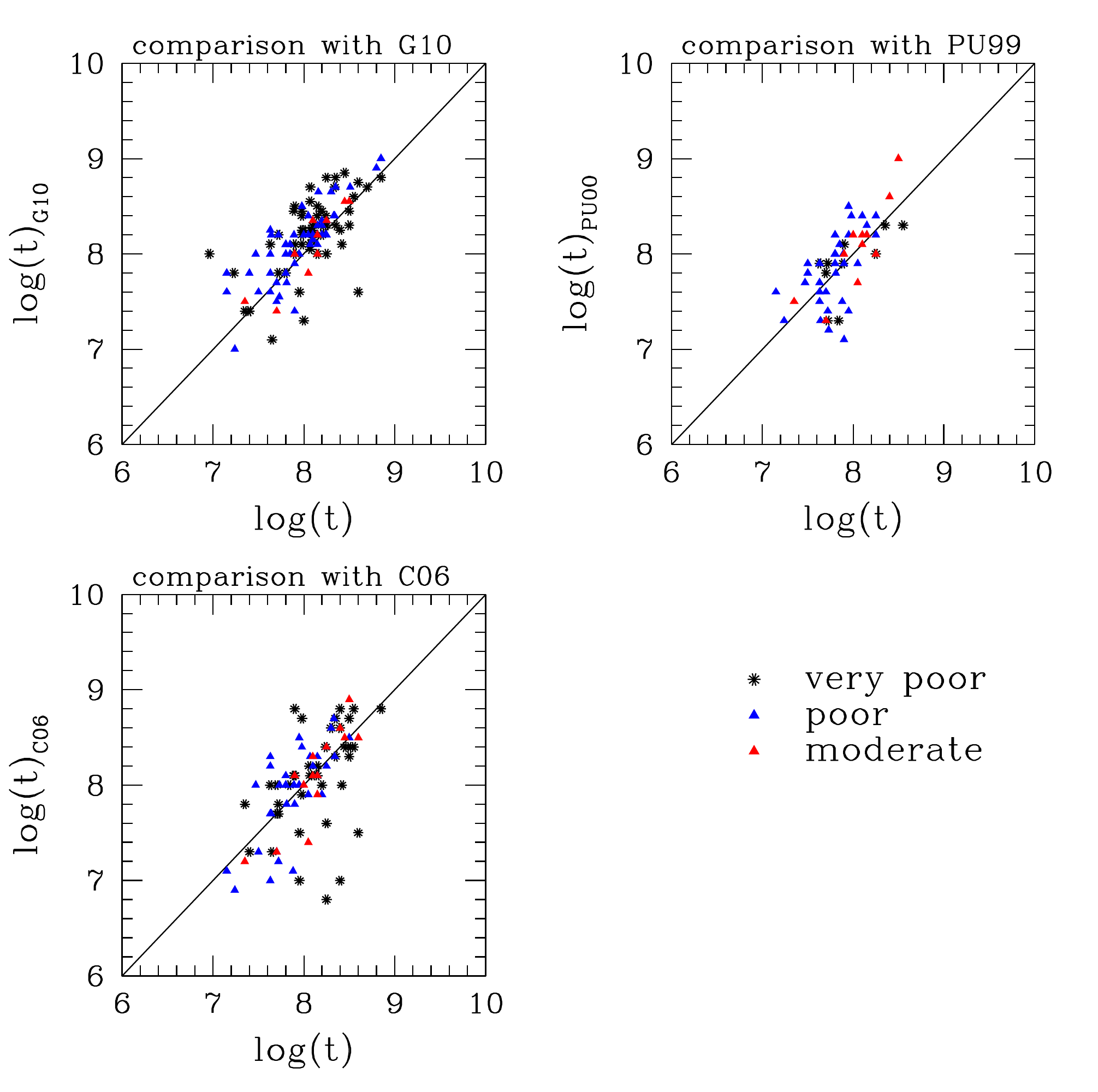} 
\caption{ Comparison between our estimated age (X-axis) and age estimated by previous studies (Y-axis). We have compared our age estimation with that of G10 (top left), PU99 (top right) and C06 (bottom one). Different point types indicate clusters from different group of classification. A straight line with slope = 1 is shown in each plot to indicate the deviation in the estimated age.}
\label{age_comp}
\end{figure*}

\section{Results and Discussion}
 	We have estimated ages and reddening of 179 star clusters in the SMC using a semi-automated quantitative method. Out of these, 17 clusters are parametrised for the first time. Out of 17 newly parameterised clusters, we find one (NGC 458) rich cluster and rest belong either to very poor or poor groups. We have also classified all the clusters based on their mass/strength for the first time. 
We have listed the results in a catalog (available online). A sample of this catalog is presented in Table \ref{catalog}. The catalog contains the name of the clusters, position (RA and Dec as given in B08), radius, estimated ages and reddening by our method, previous estimation of ages by G10, PU99, C06, and group number based on our classification. In the catalog, the clusters' name designated by an asterisk and blank spaces in columns 7 to 9 respectively imply newly parametrised clusters. There are three clusters (SK157, HW77 and HW82) in our catalog with blank spaces in columns 7 to 9, whose ages are estimated by \cite{pia2015}. 

 We have presented the field star decontaminated CMDs of all the 179 clusters, with over plotted M08 isochrones for the estimated age. Isochrones showing the typical uncertainty in the age estimation (0.25) are also over plotted. 
 The CMDs will be available only as online figures. As an example, we have presented four CMDs from four groups (I - IV) in Figure \ref{cmd}. In the figure, cluster stars are denoted as black points, red solid line denotes the isochrone corresponding to estimated age and red dashed lines denote the isochrones corresponding to the age uncertainty. The turn-off of each isochrone is marked as a red point. The name of the cluster and their corresponding age are mentioned on top of each subplot, along with their group number labeled in blue.

 \subsection{Comparison with previous studies}
	 
We found 119 clusters to be in common with G10, 56 clusters are in common with PU99 and 90 with C06. In Figure \ref{age_comp} we have compared our age estimation (X-axis) with previous estimations (Y-axis). We have drawn a straight line with slope = 1 in the plots to check the difference in age estimation. Clusters with different classification are denoted in different colours in the figure. The top left plot shows that our results match very well with G10 with an uncertainty of 0.25 in log scale except for a few very poor clusters. The top right plot also shows good matching of our results with PU99, although there are a few clusters for which we estimated older ages. Our results also match well within the error of log(t) = 0.25 with the estimation by C06 for most of the clusters older than $\sim$ 60 Myr (log(t) = 7.8) (bottom panel). In the case of clusters younger than 60 Myr, we have estimated relatively older ages. We rechecked our CMDs of those clusters where we find a discrepancy in estimated ages with previous results and reconfirmed our estimated values.

In general, the comparison indicates that our estimations compare well with the previous studies. We also note a few cases of discrepancy which can be due to reasons like different data used by different authors, the difference in the isochrone models used and the difference in adopted methods. PU99 and C06, used OGLE II data, which have lesser resolution than OGLE III data. Also, OGLE II covers the central region of the SMC, where the clusters may suffer more crowding effect. PU99 used isochrone model by \citet{bertelli1994} and C06 used the isochrone model by \cite{gira2002}. Whereas, G10 used MCPS data and two isochrone models for their analysis: Padova isochrones \citet{girardi1995} and Geneva isochrones \citep{lejeune2001} for their age estimations.

\subsection{Reddening distribution}

\begin{figure}
\includegraphics[width=\columnwidth]{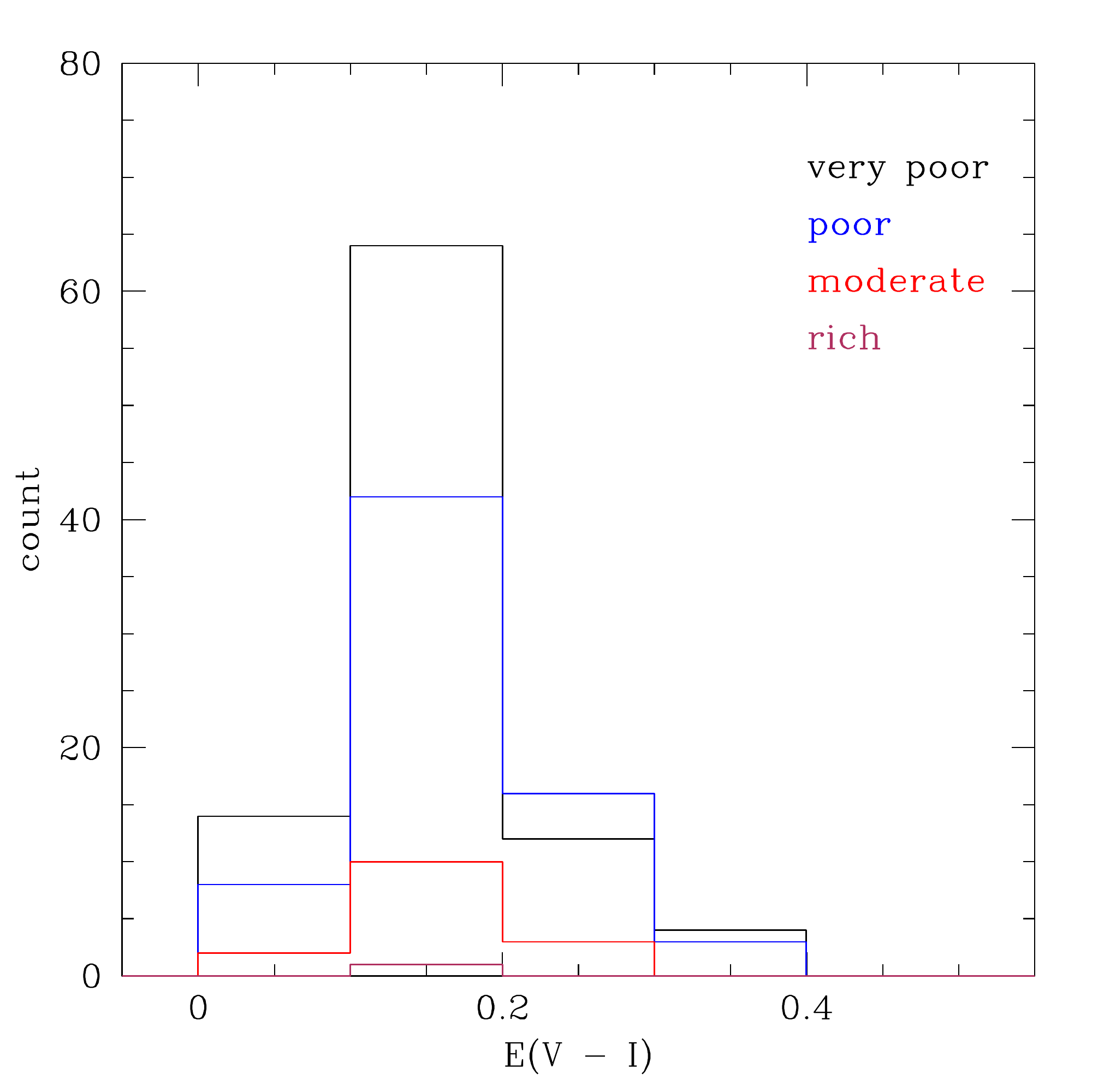} 
\caption{ Reddening distribution of very poor (black), poor (blue), moderate (red) and rich (maroon) clusters. Reddening value peaks between 0.1 and 0.2 mag for all the groups.}
\label{red_hist}
\end{figure}

		We have constructed distribution of the estimated reddening E(V $-$ I) for different groups of clusters (shown in different colour) in Figure \ref{red_hist}. 
The distribution ranges from 0 to 0.4 mag for very poor and poor clusters, and from 0 to 0.3 mag for moderately rich clusters. Whereas, the distribution  peaks between 0.1 to 0.2 mag for all the four groups. 
We have compared the estimated reddening with the high resolution map of field reddening in the central SMC by \citet{indu2011} and found that they match well. We did not find any significant difference in reddening with that estimated using red clump stars by \citet{smitha2012}.
We have also compared our estimated reddening with previous studies by G10, PU99 and C06. The distribution of difference is found to be peaked at 0.1 mag, which is within the error (1$\sigma$) of our estimation.

\begin{figure}
\includegraphics[width=\columnwidth]{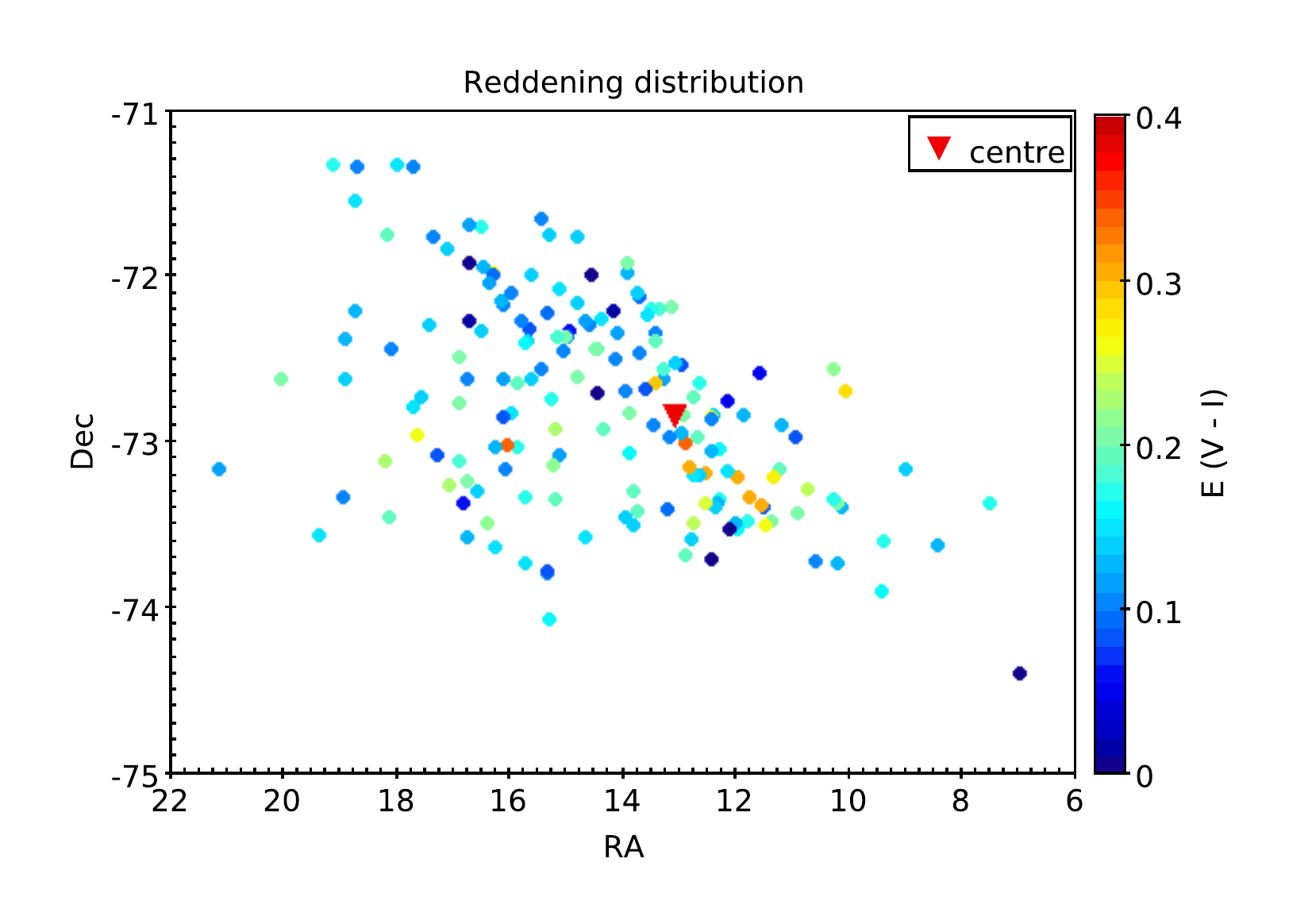} 
\caption{Spatial variation of estimated reddening across the SMC.}
\label{red_dist}
\end{figure}

		The spatial distribution of reddening across the SMC is plotted in Figure \ref{red_dist}. The red triangle indicates the centre of the SMC, located at (0 h 52 m 45 s, $-$72$\degr$ 49\textasciiacute 43\textacutedbl) \citep{crowl2001}.
Most of the regions have reddening within 0.1 - 0.2 mag with larger variation in reddening near the centre.
 The Southern part of the SMC consists of clusters with relatively large reddening value than the north eastern (NE) part.

\begin{figure}
\includegraphics[width=\columnwidth]{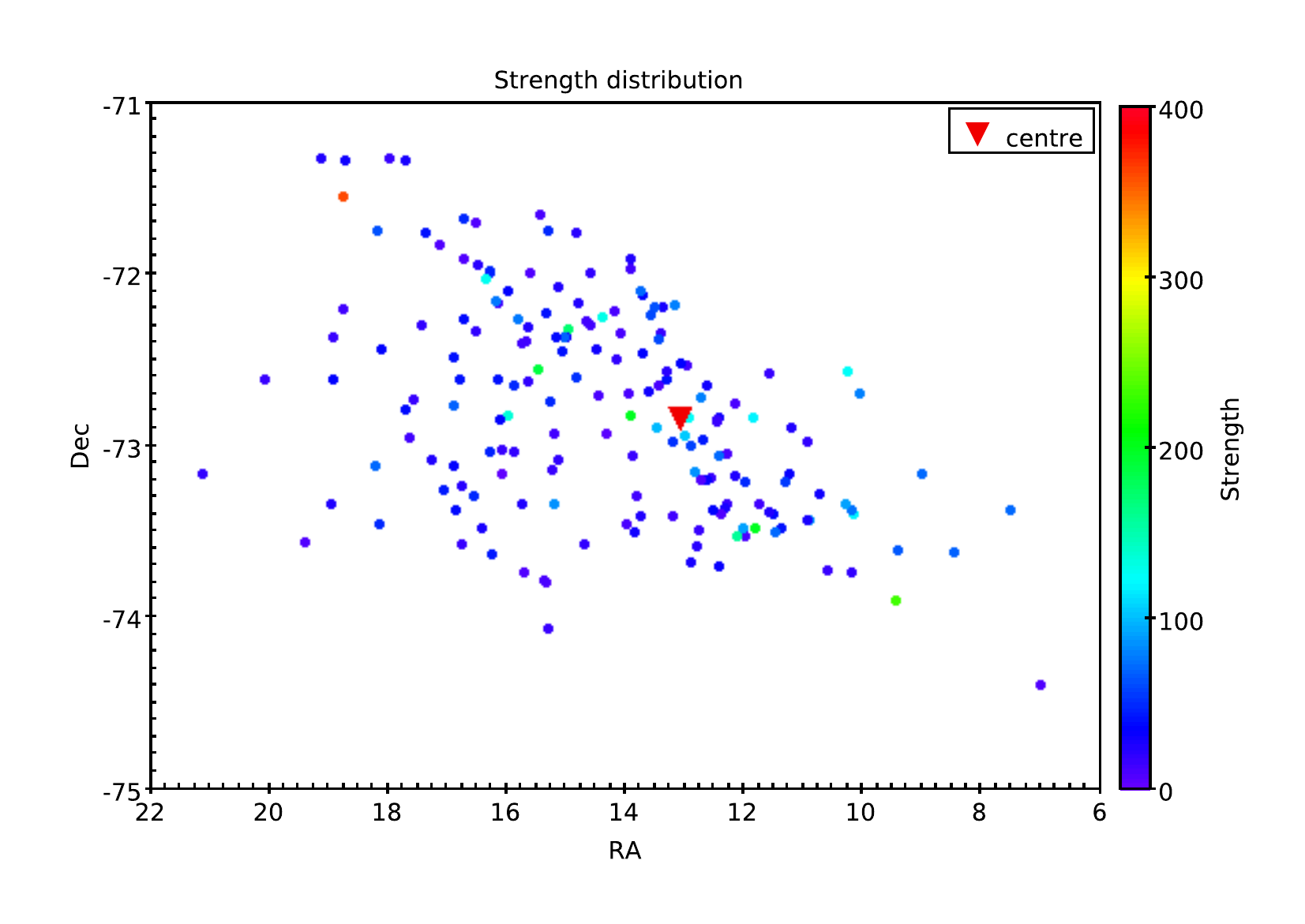} 
\caption{Spatial distribution of star clusters in the SMC as a function of cluster strength.}
\label{strength_dist}
\end{figure}

\subsection{Strength distribution}

Spatial distribution of clusters as a function of their strength ($n_m$) is shown in Figure \ref{strength_dist}. Red triangle indicates the centre of the SMC. The figure shows that clusters with $n_m <$ 100 are distributed all over the SMC observed region. The clusters with $n_m >$ 100 are preferentially located in the inner SMC, mainly close to the bar. We notice that there are two clusters with strength more than 200, located in the south western (SW) and NE end of the SMC. We do not find any kind of hierarchical distribution of clusters in the SMC based on their strength/mass, similar to that found in the LMC (Paper I).  
We notice the presence of only low mass clusters in the eastern SMC which is predominantly affected by the tidal forces \citep{besla2010,besla2012}.

\begin{figure}
\includegraphics[width=\columnwidth]{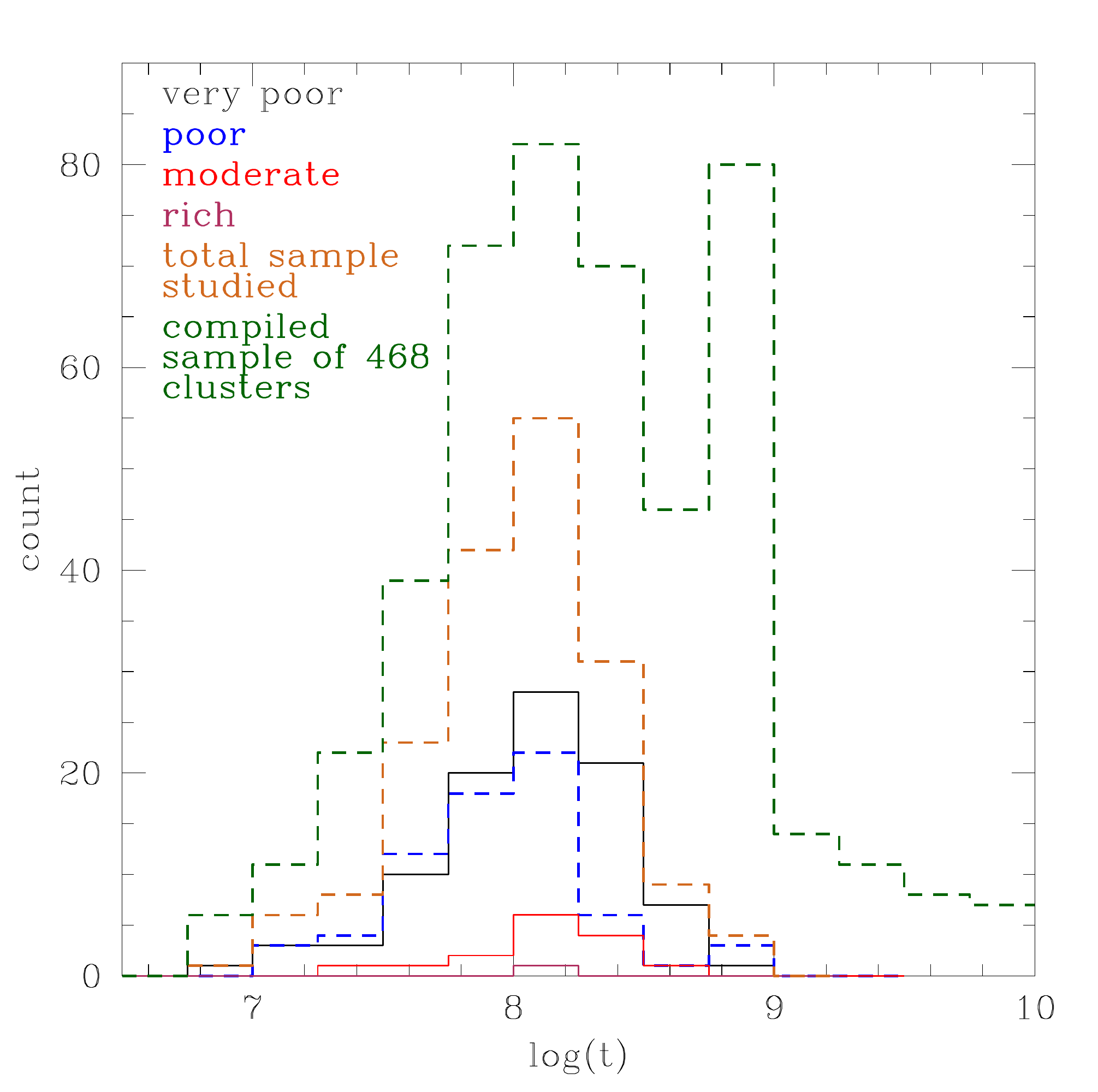} 
\caption{Age distribution of very poor (black), poor (blue), moderate (red) and rich (maroon) clusters. Distributions of all the four groups of clusters peak in the range of log(t)=8.00-8.25. Cumulative age distribution of all the studied clusters (chocolate) shows a peak at $\sim$ 130 $\pm$ 35 Myr. Age distribution of the compiled sample of 468 clusters peaks at $\sim$ 130 and $\sim$ 750 Myr.}
\label{age_hist}
\end{figure}

\begin{figure}
\includegraphics[width=\columnwidth]{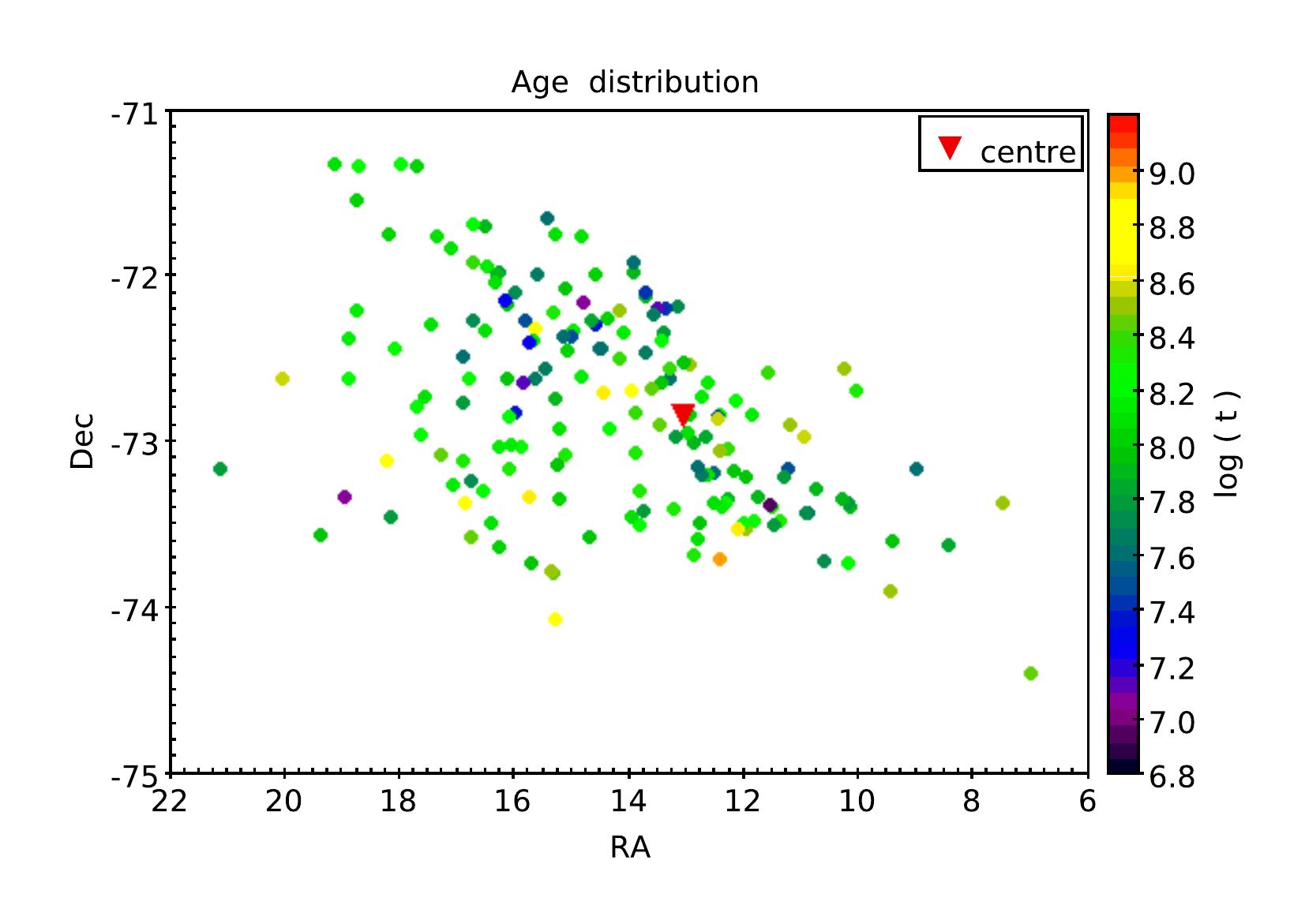} 
\caption{Spatial distribution of star clusters in the SMC as a function of age.}
\label{age_dist}
\end{figure}

\begin{figure}
\includegraphics[width=\columnwidth]{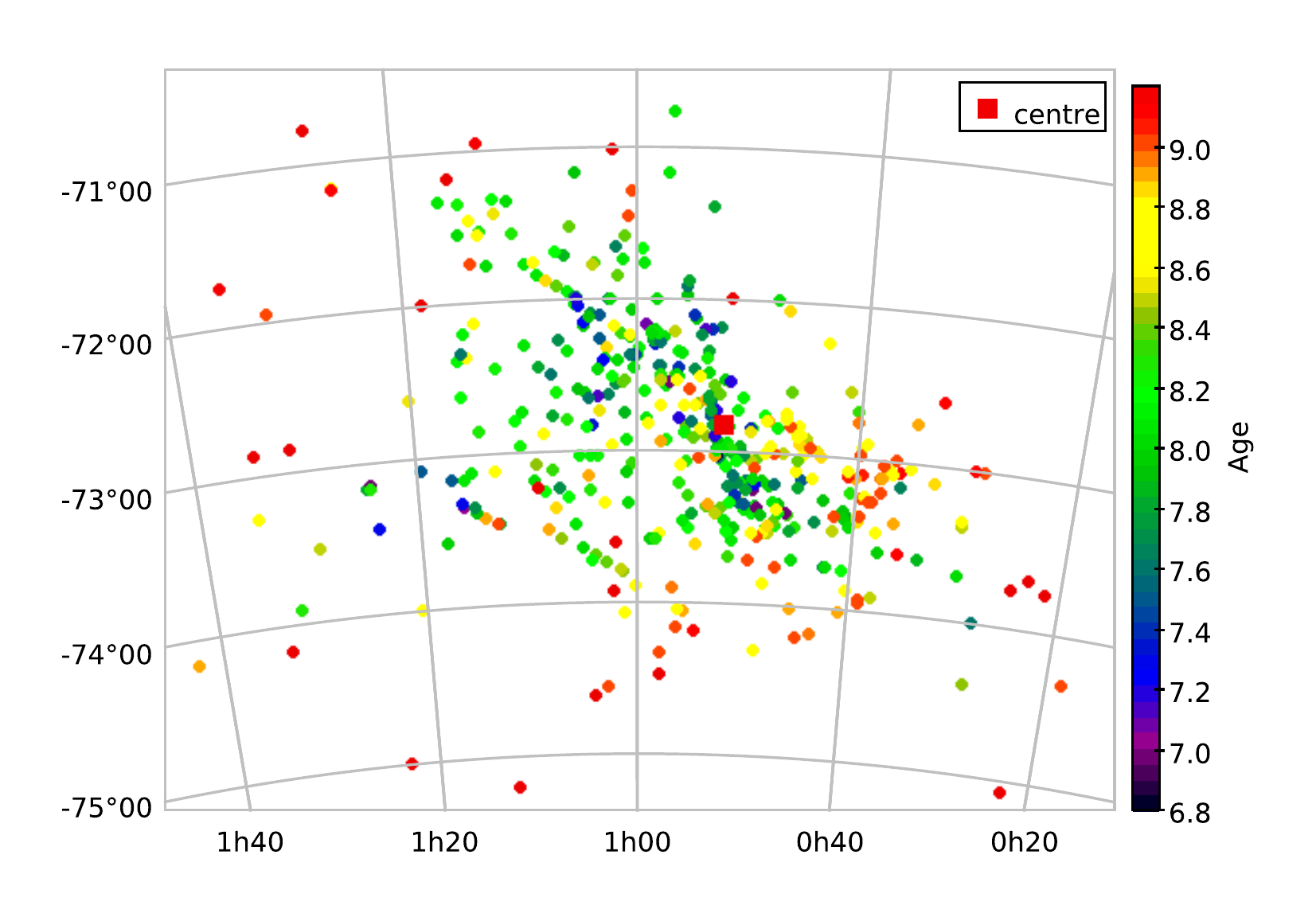} 
\caption{Spatial distribution of the compiled sample of 468 star clusters in the SMC as a function of age.}
\label{compiled_sample}
\end{figure}

\begin{figure*}
\centering
\includegraphics[width=1.45\columnwidth]{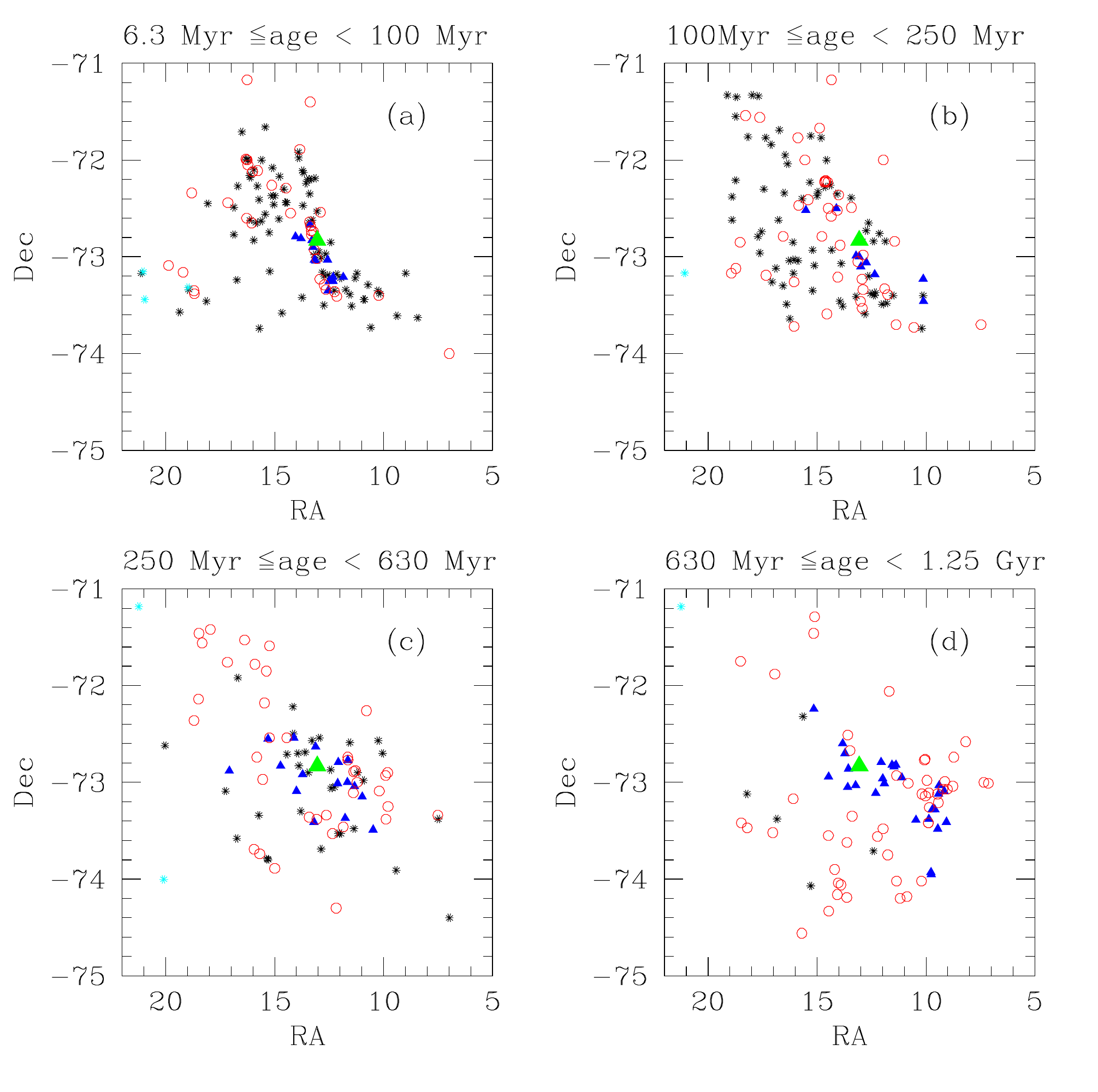} 
\caption{ Plot shows spatial location of the SMC clusters at different epochs in each panel, where the clusters are taken from our catalog (black), by G10 (red circle), C06 and PU99 (blue), $'$other-studies$'$ (cyan).}
\label{age_range}
\end{figure*}

\subsection{Age distribution}
		Age distribution of various groups of clusters is shown in Figure \ref{age_hist}. We have used a bin width of 0.25 in log scale which is same as the error associated with the age estimation. The figure shows that very poor clusters are distributed over a large age range (log(t) = 6.75 - 9.00 ) with peak at $\sim$130 Myr (log(t) = 8.00-8.25).  The poor clusters show two peaks : a younger peak at 130 Myr (log(t) = 8.00-8.25) and an older peak at 750 Myr (log(t) = 8.75-9.00). 
		The moderately rich and rich clusters also show peak at 130 Myr (log(t) = 8.00-8.25), similar to the very poor and poor clusters. 
Therefore, most of the clusters are mainly distributed between 30 Myr to 300 Myr (log(t) = 7.5 - 8.5). The cumulative distribution of all the studied clusters peaks at $\sim$ 130 Myr, which is almost same as that we identified in the LMC (Paper I). The fact that the cluster formation peaked at $\sim$ 130 Myr in both the MCs could suggest that it is due a common triggering event. We suggest that the recent interaction ( $\sim$ 200-300 Myr ago) between the LMC and SMC might have triggered the cluster formation during the above mentioned age range. PU99 found the peak of cluster formation at around 30 Myr, whereas C06 found two peaks of cluster formation at 8 Myr and 90 Myr.   
G10 found two peaks of cluster formation at 160 and 630 Myr. The younger peak of G10 is not very different from our younger peak.

The spatial distribution of age is shown in Figure \ref{age_dist}. The figure shows that clusters with age around 100 Myr are distributed all over the SMC region. Western and south eastern (SE) parts show clusters with older ages. The central part consists of clusters with a relatively larger age range. 
The SMC stretches out from SW to NE direction and younger clusters are found in specific places, slightly north from the central region.
Older clusters are mostly located in southern part of the SMC and northern part consists of relatively younger clusters. 
This is suggestive of preferential location of clusters as a function of age. These needs to be confirmed with a more complete sample of parameterized clusters.

 We have therefore added 289 clusters from the previous studies which are not common to our catalog. The clusters and their parameters are taken from G10, PU99, C06, \cite{pia2005,pia2007b,pia2007a,pia2007c,pia2008,pia2011a,pia2011b,maia2012,pia2012a,mighell1998,glatt2008,girardi2013,crowl2001,parisi2014}, \cite{crowl2001,dias2014,pia2015}. As there are a large number of studies, we use a common reference, $'$other-studies$'$, to indicate all the studies other than G10, PU99 and C06. The addition of clusters from previous studies not only increased the cluster sample but also the coverage of the SMC. The total sample of 468 clusters is large enough to study the spatio-temporal distribution of clusters, which is discussed in the next section. This is the largest parameterised sample of clusters in the SMC.
The age distribution of this compiled sample of 468 clusters is also shown in Figure \ref{age_hist}. The distribution shows two peaks (130 and 750 Myr) of cluster formation. The younger peak is found to be same as that estimated from our sample.

Spatial distribution of the compiled sample as a function of age is shown in Figure \ref{compiled_sample}. The centre of the SMC is denoted by solid red square. The distribution suggests that older clusters are mostly located at the southern and western part of the SMC, whereas, the younger clusters are found in the inner SMC along with a few clusters in the east. The clusters with ages $\sim$ 100 Myr are distributed all over the SMC. The distribution also suggests that the SMC is stretched out from SW to NE along the bar, could be due to the interaction between the MCs. We find that the north west quadrant of the Fig. \ref{compiled_sample} is devoid of clusters. We suggest that the reasons could be either due to lack of available photometric data in that region resulting in no parameterised clusters, or due to genuine lack of clusters in this part of the SMC. It is important to fill this gap in the spatial distribution of SMC clusters. A genuine lack of clusters in this region can put constraints on the cluster formation as well as gas distribution in the SMC.

\subsection{Spatio-temporal distribution}

 To understand the spatio-temporal distribution, the spatial location of clusters in various age range is shown in Figure \ref{age_range}.  
The ages of the clusters are in the range of log(t) = 6.8 to 9.1. The black points in the figure denote the clusters from our catalog, the red small circles are the clusters from G10, the blue points are from C06 and PU99, and the clusters from $'$other-studies$'$ are denoted as cyan points. The green point indicates the centre of the SMC.
As shown in Figure \ref{age_range}(d) clusters in the age range 630 Myr - 1.25 Gyr are mostly found in the southern and western parts of the SMC including the central region. Very few clusters are found in the northern and eastern regions during this period. During the period 250 - 630 Myr (Figure \ref{age_range}(c)), the clusters are found mostly in the central region, along with a group of clusters in the NE region. 
On the other hand, the western and the southern regions are devoid of clusters. 

In the age range 100 - 250 Myr (Figure \ref{age_range}(b)), most of the clusters are found in the eastern and NE regions along with the central SMC. The western and southern regions continue to be devoid of clusters during this period. We also find that the extent of the NE region is maximum during this period. Figure \ref{age_range}(a) shows the location of clusters formed in the last 100 Myr. These are found to be mostly in the NE region and the central SMC. We notice a specific pattern in the distribution of clusters, which is different from the other three panels.

Fig.\ref{age_range}(a) and (d) show the distribution of clusters in two extreme epochs. The spatial distribution can be seen to be distinctly different with no co-relation between the two epochs. Most of the clusters in the older epoch are in the southern part, whereas the ones in the younger epoch are mostly found in the northern and central region. The panels Fig.\ref{age_range}(b) and (c) show the shift of clusters from south to north. 
Similar shift was found in star formation during the same period by \cite{harris2004} (their Fig.6). Figure \ref{age_range} also suggests that the central region of the SMC is actively forming clusters from $\sim$ 1 Gyr till date. 
We suggest that a close interaction between the LMC and the SMC 1.2 Gyr ago (\citet{diaz2011}) may be the reason for triggering cluster formation in the southern and the western part of the SMC (Figure \ref{age_range}(d)). We also suggest that the recent interaction at 250 Myr caused cluster formation in the last 100 Myr, resulting in the spatial distribution as shown in Figure \ref{age_range}(a). Figure 13 of \citet{besla2012} showed that the SMC made close passages around the LMC at $\sim$ 900 Myr and 100 Myr ago, which supports the above observation. The spatial distribution of clusters presented in Fig.\ref{age_range} could give important clues regarding the details of the interactions.

		We made two videos (available online only) to understand the spatio-temporal distribution in detail. 
		In the video-1, we have shown the cluster distribution from older to younger age, and video-2 shows vice-versa. In the videos, we have used the same color notations as in Figure \ref{age_range}. The two videos clearly demonstrate the change in location of clusters as a function of age. The details of spatio-temporal distribution of this largest cluster sample will provide important details of cluster formation history in the SMC. The distribution shown in Figure \ref{age_range} are in fact snapshots from the videos for specific epochs. Many such snapshots can be created for various epochs as required using these videos.

		In the case of the LMC (Paper I), we identified an outside to inside propagation of cluster formation. On the other hand, in the SMC, we identify a progressive shifting of cluster location from the South to the North during the last 600 Myr. The clusters older than 1.25 Gyr are found to be distributed in the outskirts of the SMC. We identified both the MCs to have a peak in cluster formation at $\sim$ 130 Myr. This is suggestive of a common cluster formation trigger, which is most likely to be the recent interaction between the MCs. The details of spatio-temporal distribution of clusters presented in this study together with Paper I can be used as a tool to constrain details of the recent LMC-SMC interactions.

\begin{figure}
\includegraphics[width=\columnwidth]{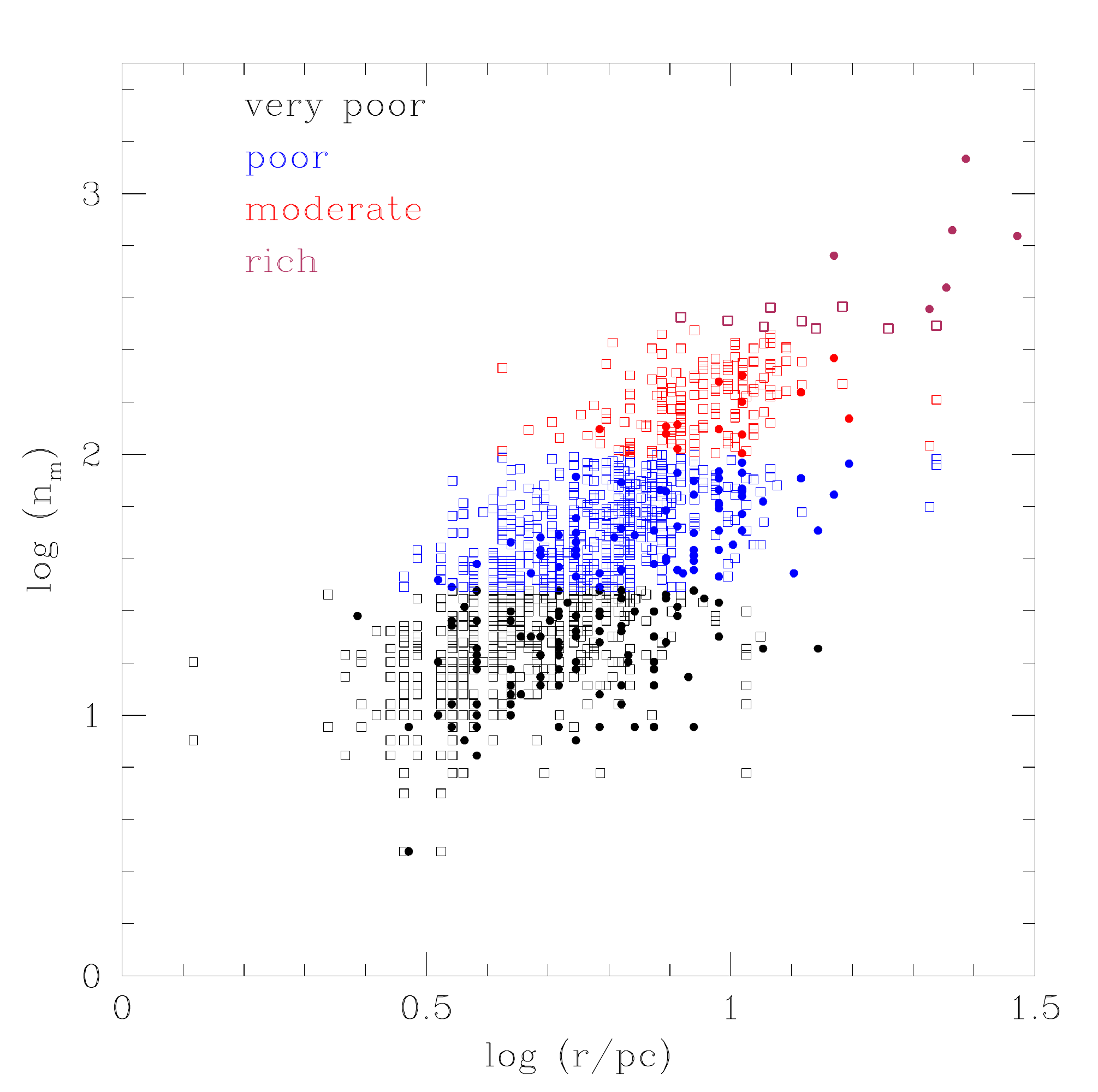}
\caption{Plot shows the relation between radius and strength of clusters in the LMC (open box) and the SMC (filled circle). Clusters of different classification are denoted by different colors.}
\label{mass_nm}
\end{figure}

\begin{figure}
\includegraphics[width=\columnwidth]{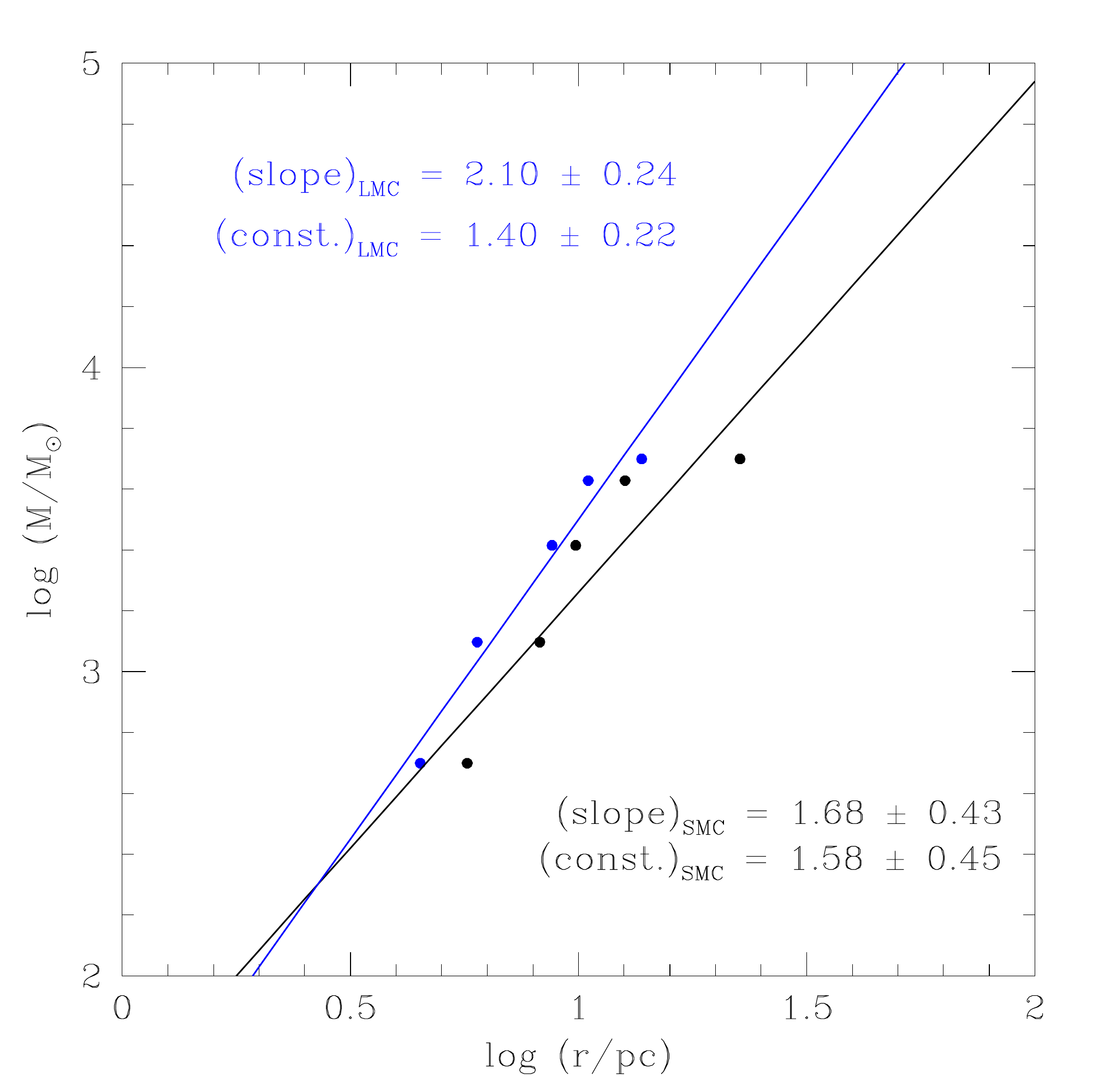} 
\caption{The relation between average radius and average mass of clusters of different groups. The blue data points correspond to the LMC and the black ones correspond to the SMC. The slope and y-intercept of linear fit are also mentioned in the Figure.}
\label{mass_radius}
\end{figure}

\subsection{Mass-Radius relation}

In order to understand the structure of the SMC clusters, we have plotted radius of clusters (log(r)) against the strength (log($n_m$)) for our studied sample (filled circle) in Figure \ref{mass_nm}. There are five clusters with $n_m > 400$ (categorised as rich clusters), which we excluded for parameterisation, are also shown in this figure. To compare the structure of the SMC clusters with that of the LMC, we have over plotted the LMC clusters data (open box), taken from Paper I. Clusters with different classification are denoted in different color. The figure shows that there is linear relation between radius and strength in logarithmic scale for both the MCs. The figure also suggests that though there is a spread in the radius of clusters with similar strength for both the MCs, the SMC clusters tend to have systematically larger radii than LMC clusters with similar strength. This points to the possibility of compactness of the LMC clusters  when compared to the SMC clusters. In order to shed more light on this, it is necessary to estimate the nature of relation between cluster mass and radius. 

Figure \ref{mass_radius} shows the relation between radius (log(r)) and mass (log M) of clusters in the SMC. 
We took an average of mass range of clusters and average radius of clusters for different groups (I - V) to estimate the co-relation between them. The data points of the LMC and SMC are marked as blue and black respectively. Straight lines fitted to the data points give slopes of 2.10 ($\pm$0.24) and 1.68 ($\pm$0.43) for the LMC and the SMC respectively. The difference in slope if of the oder of 1-$\sigma$, hence the result is only indicative. There is an indication that the clusters with similar mass occupy smaller radius in the case of the LMC than in the SMC, indicating that the SMC clusters are loosely bound when compared to those in the LMC. 
 \citet{pfal2016}  derived a similar relation for star clusters in the solar neighbourhood using the relation M$_c$ =C${_m}{\times}R^{\gamma}$. They found the value of $\gamma$ as 1.7$\pm$0.2 for a large range of cluster mass. They also mentioned that it is necessary to find out if there exists a universal relation between mass and radius of clusters.  
  We find that the values of $\gamma$ are similar for the SMC and the Galaxy, whereas it is marginally higher for the LMC. This suggests that cluster formation environment in the SMC is similar to that of our Galaxy in the solar neighbourhood. Our study shows that there is a tentative evidence for tighter clusters in the LMC, when compared to those in the Galaxy and the SMC. This needs to be verified with the help of better data with individual mass and radii estimates of clusters.

\section{Summary}

1. We have estimated the age and reddening of 179 star clusters in the SMC using OGLE III data and presented a catalog (available on-line). Out of 179, 17 clusters are parameterised for the first time. Out of 17 newly parameterised clusters, one is rich (NGC 458) while the rest belong to very poor or poor group.
 \\
2. Field star decontaminated CMDs of all the 179 clusters, fitted with isochrones of estimated age and corrected for reddening are available online. \\
3. We have also classified the SMC star clusters based on their mass and richness in four groups for the first time. \\
4. We find that 90 $\%$ of our studied sample has mass $<$ 1700 M$_\odot$, which suggests that the SMC is dominated by low mass clusters. The lower mass limit of the SMC star clusters is found to be very similar to that of the open clusters in the Galaxy. We also find a tentative evidence for tighter clusters in the LMC, when compared to the LMC and our Galaxy.\\
5. Combining our sample with previous studies, we compiled age information of 468 clusters to study their spatio-temporal distribution.
We find the age distribution to peak at 130 $\pm$ 35 Myr, similar to the LMC (Paper I). We suggest that this could be due to the most recent LMC-SMC interaction. \\ 
6. The clusters with age 630 Myr - 1.25 Gyr are found to be located preferentially in the South and West of the SMC, whereas the clusters younger than 100 Myr are found in the North and eastern regions, suggesting a shift in the location of cluster formation. The central SMC shows a continuous formation of clusters in the last 1 Gyr. The details of this spatio-temporal shift is presented in two videos (available on-line). \\
7. The details of spatio-temporal distribution of clusters presented in this study together with Paper I can be used as a tool to constrain details of the recent LMC-SMC interactions. \\

\section{ACKNOWLEDGEMENTS}
	Ram Sagar would like to acknowledge the award of NASI-Senior Scientist Platinum Jubilee Fellowship by the National Academy of Science, Allahabad, India. S. Choudhury would like to thank the support from Basic Science Research Program through the National Research Foundation of Korea (NRF) funded by the Ministry of Education (NRF2016R1D1A1B01006608), and that by the KASI-Yonsei Joint Research Program for all Frontiers of Astronomy and Space Science funded by the Korea Astronomy and Space Science Institute. P. K. Nayak would like to thank Dr. Avijeet Prasad (Udaipur Solar Observatory, Physical Research Laboratory, India) for helping with Mathematica code to make the videos. The authors thank the OGLE team for making the data available in public domain.

\begin{table*}
\centering
\caption{A sample of the complete catalog is presented. The table contains the name of the cluster, central coordinates (RA and Dec) as given in B08, size of the cluster taken from B08,  estimated reddening and age, in columns 1-6 respectively. Columns 7-9 contain the earlier estimations of ages by G10 (log(t$_{G10}$)), PU99 (log(t$_{PU99}$)) and (log(t$_{C06}$)). The last column contains the designated group number (I-V).}
\label{catalog}
\begin{tabular}{lccccccccc}
\hline
Star cluster & Ra & DEC & Radius & E(V$-$I) & log(t) & log(t$_{G10}$) & log(t$_{PU99}$) & log(t$_{C06}$)  & Group \\
 & (h m s) & ($\degr$ $\arcmin$ $\arcsec$) & ( $\arcmin$ ) & & & & & & \\
\hline
\hline
B6*             &  0 27 57 &  -74 24 02 &  0.30 &  0.10 &  8.25 &     $-$ &    $-$ &   $-$ &    I \\
K9              &  0 30 00 &  -73 22 45 &  0.60 &  0.17 &  8.60 &    8.70 &    $-$ &   $-$ &   II \\
HW8             &  0 33 46 &  -73 37 59 &  0.85 &  0.11 &  7.90 &    8.00 &    $-$ &   $-$ &   II \\
NGC176          &  0 35 58 &  -73 09 58 &  0.60 &  0.12 &  7.64 &    8.20 &    $-$ &   $-$ &   II \\
HW11            &  0 37 33 &  -73 36 43 &  0.65 &  0.15 &  8.20 &    8.50 &    8.4 &   8.4 &   II \\
L19             &  0 37 42 &  -73 54 30 &  0.85 &  0.16 &  8.70 &     $-$ & $>$9.0 &   8.9 &   IV \\
B14             &  0 38 37 &  -73 48 21 &  0.26 &  0.10 &  8.25 &    8.65 &    $-$ &   7.9 &    I \\
HW12            &  0 38 51 &  -73 22 27 &  0.40 &  0.09 &  8.45 &    8.70 &    $-$ &   8.7 &    I \\
H86-48          &  0 38 56 &  -73 24 32 &  0.22 &  0.07 &  8.30 &     $-$ &    $-$ &   8.0 &    I \\
SOGLE6          &  0 39 33 &  -73 10 37 &  0.40 &  0.17 &  8.70 &    8.65 &    $-$ &   $-$ &    I \\

\hline
\end{tabular}
\end{table*}

\bibliographystyle{aa}
\bibliography{ref_paper2}

\end{document}